\documentclass{iaus}
\usepackage{graphicx}

\title[Structure and evolution of star-forming gas in late-type spiral galaxies] 
{Structure and evolution of star-forming gas in late-type spiral galaxies}

\author[Fathi et al.]   
{Kambiz Fathi$^{1,2}$, 
John E. Beckman$^{1,3}$, 
Almudena Zurita$^4$, 
M\'onica Rela\~no$^4$, 
Johan H. Knapen$^{1,5}$, 
G\"oran \"Ostlin$^2$, 
Claude Carignan$^6$, 
Laurent Chemin$^7$, 
Olivier Daigle$^6$, 
Olivier Hernandez$^6$}

\affiliation{
$^1$Instituto de Astrof\'\i sica de Canarias, C/ V\'\i a L\'actea s/n, 38200 La Laguna, Tenerife, Spain \break email: fathi@iac.es\\[\affilskip]
$^2$Stockholm Observatory, AlbaNova University Center, 106 91 Stockholm, Sweden\\[\affilskip]
$^3$Consejo Superior de Investigaciones Cient\'\i ficas (SCIC), Spain\\[\affilskip]
$^4$Depto. de F\'{\i}sica Te\'orica y del Cosmos, Universidad de Granada, 18071 Granada, Spain\\[\affilskip]
$^5$Centre for Astrophysics Research, University of Hertfordshire, Hatfield, Herts AL10 9AB, UK\\[\affilskip]
$^6$Universit\'e de Montr\'eal, C.P. 6128 succ. centre ville, Montr\'eal, QC, Canada H3C 3J7\\[\affilskip]
$^7$Observatoire de Paris, Section Meudon, GEPI, CNRS-UMR 8111, Meudon, France
}

\pubyear{2007}
\volume{241}  
\pagerange{119--126}
\date{?? and in revised form ??}
\jname{Stellar Populations as Building Blocks of Galaxies}
\editors{Peletier R.F. \& Vazdekis A., eds.}
\begin{document}

\maketitle

\begin{abstract}
We study two dimensional Fabry-Perot interferometric observations of the nearby 
face-on late-type spiral galaxy, NGC~628. We investigate the role of the individual {\sc H$\,$ii} regions 
together with the large-scale gravitational mechanisms which govern star formation and overall 
evolution in spiral galaxies.  Our kinematical analysis (reinforced by literature maps in HI and CO at
lower angular resolution) enables us to verify the presence of an inner rapidly rotating inner disk-like component 
which we attribute to long term secular evolution of the large-scale spiral arms and oval structure.  
We find that gas is falling in from the outer parts towards the bluer central regions. 
This could be an early phase in the formation of a pseudo-bulge. We find signatures of 
radial motions caused by an $m=2$ perturbation, which are likely to be responsible for the inflow of 
material forming the circumnuclear ring and the rapidly rotating inner structure.

\end{abstract}

\firstsection 

\section{Introduction}
The line-of-sight velocity and velocity dispersion ($\sigma$) are important parameters in determining the flattening of 
the various disc layers, the intrinsic shape of the dark matter component, and the nature and extent of disc-halo 
interactions. If the gaseous layers in the disc are isothermal, 
the atomic and molecular gas should have similar scale-heights. However, due to the collective effects of 
gravitational instabilities, dissipation, and feedback from star-forming regions, the atomic gas has been observed 
to have a distribution different from that of its molecular counterparts. 

\section{Our Programme}
We have launched an extensive effort to analyse the processes that govern the ionized gas 
and compare them with those that control the neutral gaseous components in spiral galaxies 
(see \cite[Fathi et al. 2007]{Fathietal2007}). We use Fabry-Perot interferometry with instrumentation presented in 
\cite[Hernandez et al. (2003)]{Hernandezetal2003} and scan the H$\alpha$ emission-line. Our observations 
yield the distribution and kinematics of the H$\alpha$-emitting gas. Quantifying the kinematic parameters, and 
comparing with those of the neutral gas will constrain the dynamics of the star-forming regions.

\begin{figure}
 \includegraphics[width=.99\textwidth]{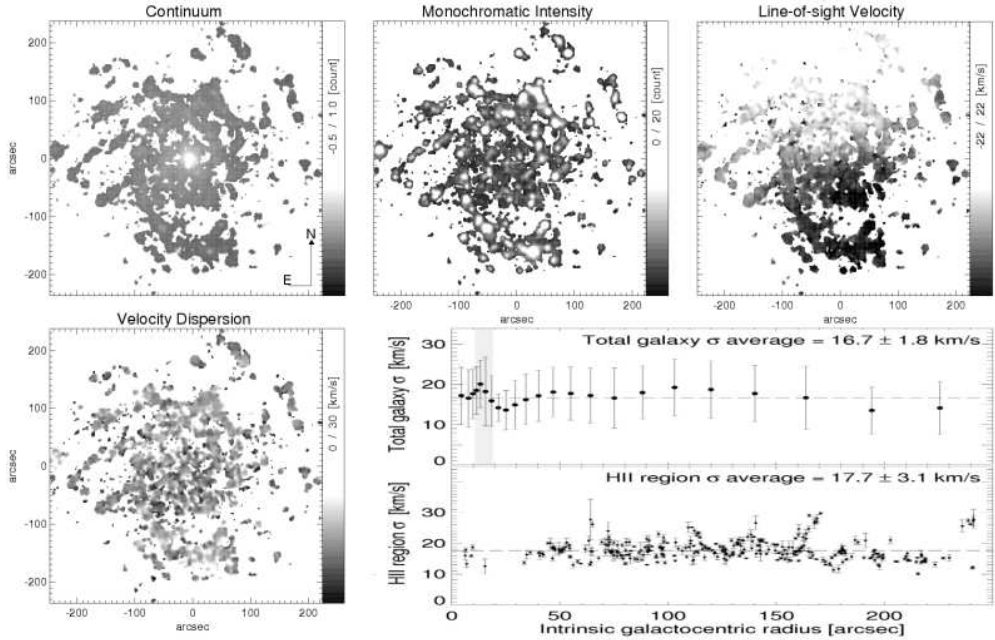}
  \caption{The distribution and kinematics of the H$\alpha$ emitting gas in NGC~628 (top), and the velocity
   dispersion map with its galactocentric profile (bottom). The velocity field has been quantified 
   using the harmonic decomposition technique, and the $\sigma$ of the individual {\sc H$\,$ii} regions 
   has been derived after using our {\sc H$\,$ii} region catalogue to identify them. (see \cite[Fathi et al. 2007]{Fathietal2007}).} 
  \label{fig:results}
\end{figure}

We have chosen NGC~628 for a pilot study in which we demonstrate the power of our state-of-the-art methods 
to study the properties of the ionized gas (mainly {\sc H$\,$ii} regions but also diffuse ionized gas) and their 
relation with the global kinematic parameters (e.g., \cite[Zurita et al. 2004]{Zuritaetal2004}, 
\cite[Fathi et al. 2005]{Fathietal2005}). A key feature of the velocity field of NGC~628 is its regularity. 
The global effects of any asymmetries, such as the oval distortion, are small, and in any case not very 
easy to detect in a face-on system. However, our detailed kinematic analysis has revealed the presence 
of a disc-like component in the inner kpc around the nucleus. The $\sigma$ map shows widely 
distributed star-formation in the disc plane, and that the emission from unresolved {\sc H$\,$ii} 
regions probably dominates any emission from the diffuse component. The radial $\sigma$-profile shows  
a nearly constant value of $\approx17$ km/s (more than twice that for the CO and {\sc H$\,$i}) 
out to 12 kiloparsec. Our findings for NGC~628 are presented in Fig.~\ref{fig:results}, 
and summarized in the abstract of this brief article.

\begin{acknowledgments}
We thank the secretarial staff of the IAC, the LOC, and the SOC for an enjoyable and 
stimulating conference.
\end{acknowledgments}

\end{document}